\RequirePackage{fix-cm}
\documentclass[twocolumn]{svjour3}          
\smartqed  

\usepackage{amsmath,amssymb,amsfonts,mathtools}
\usepackage{textcomp}
\usepackage{graphicx}
\usepackage{psfrag}
\usepackage{color,soul}
\usepackage{float}
\usepackage{cite}
\usepackage{authblk}
\usepackage{array}
\usepackage{url}
\newcommand\Mark[1]{\textsuperscript#1}

\DeclarePairedDelimiter\abs{\lvert}{\rvert}

\begin{document}

\title{Survey on Physical Layer Security for 5G Wireless Networks 
}

\author{\textbf{Jos\'e David Vega S\'anchez}\Mark{1} \and \textbf{Luis Urquiza-Aguiar}\Mark{1} \and \textbf{Martha Cecilia Paredes Paredes}\Mark{1} \and \textbf{Diana Pamela Moya Osorio}\Mark{2}}



\authorrunning{$^{ }$} 

\institute{ 
Jos\'e David Vega S\'anchez\at
              jose.vega01@ epn.edu.ec \\ \\
              Luis Urquiza-Aguiar\\
              luis.urquiza@epn.edu.ec\\ \\
              Martha Cecilia Paredes Paredes   \\
              cecilia.paredes@epn.edu.ec\\ \\  
               Diana Pamela Moya Osorio  \\
           diana.moyaosorio@oulu.fi\\ \\     
$^1$Departamento de Electr\'onica, Telecomunicaciones y \\ $^{ }$ Redes  de Informaci\'on, Escuela Polit\'ecnica Nacional (EPN), \\ $^{ }$ Quito, 170525, Ecuador\\ $^2$Centre for Wireless Communications (CWC), University of\\ $^{ }$  Oulu, Oulu, 90014, Finland
 }

\date{Received: date / Accepted: date}

\maketitle

\begin{abstract}
Physical layer security is a promising approach that can benefit traditional encryption methods. 

The idea of physical layer security is to take advantage of the features of the propagation medium and its impairments to ensure secure communication in the physical layer. This work introduces a comprehensive review of the main information-theoretic metrics used to measure the secrecy performance in physical layer security. Furthermore, a theoretical framework related to the most commonly used physical layer security techniques to improve the secrecy performance is provided. Finally, our work surveys physical layer security research over several enabling 5G technologies, such as massive multiple-input multiple-output, millimeter wave communications, heterogeneous networks, non-orthogonal multiple access, and full-duplex. Also, we include the key concepts of each of the aforementioned technologies. 
Future fields of research
 and technical challenges of physical layer security are also identified.

\end{abstract}

\section{Introduction}
\label{intro}
The increasing demands for wireless applications and the rapid growth of the number of connected users have saturated the capacity of current wireless communication systems. These fundamental problems motivate researchers and network designers to provide novel solutions that guarantee ultra-high data rates, ultra-wide radio coverage, a massive number of efficiently connected devices, ultra-low latency, and efficient energy consumption. In this sense, the fifth generation of wireless networks (5G) foresees great advances on solutions that satisfy these stringent requirements by employing intelligent and efficient technologies~\cite{What}. Accordingly, 5G must be prepared to tackle major challenges concerning the reliability, security, and efficiency of the network. Specifically, the security paradigm protecting the confidentiality of wireless communication is one of the core problems to be considered in 5G~\cite{Bisson}.Unlike the traditional security systems that are based on higher layer cryptographic mechanisms~\cite{Stallings}, physical layer security (PLS) uses the inherent randomness (e.g., noise and fading) of the wireless channel to ensure secure communications in the physical layer~\cite{Bisson}. In particular, PLS offers a great advantage comparing cryptographic algorithms, since it does not rely on computational complexity. Therefore, the security level achieved will not be affected even if the eavesdropper has unlimited computing capabilities. This contrasts with encryption-based approaches, which is based on the idea that eavesdropper has reduced computational capabilities to solve difficult mathematical problems in limited periods~\cite{Elkashlan}.

The first ideas of PLS are from the seminal paper of Shannon, who laid the basis of secrecy systems~\cite{Shannon}. Later, the wiretap channel was presented by Wyner in 1975~\cite{Wyner}. In that work, Wyner established that secret messages can be transmitted when the wiretap channel is a degraded (much noisier) version of the legitimate link. Thus, the secrecy capacity is the maximum data rate that can be safely transmitted without being decoded by an eavesdropper. In practice, due to the intrinsic randomness of the medium, the signal-to-noise ratio (SNR) of the eavesdropper can be similar or even better than the legitimate channel. Specifically, when the eavesdropper is closer to the source than the legitimate receiver. So, Wyner's ideas become impracticable in such environments. Inspired by Wyner's work, investigations of the attainable secrecy capacity against eavesdropping were addressed in~\cite{Korner} for the broadcast channel, and the Gaussian channel in~\cite{Yan}. These approaches have inspired an important amount of recent research activities from the information theoretic point of view for different fading channels. 
Specifically, we survey the fading channel models that have proven to accurately characterize mm-Wave scenarios in 5G. We can mention the following: $1)$ $\kappa$-$\mu$ shadowed: In this fading model, the received power signal is structured by a finite sum of multipath clusters. Each cluster is modeled by a dominant component and scattered diffuse waves. All the specular components are subject to the shadowing fluctuation caused by obstacles or human body movements~\cite{Pariss}. The PLS performance over $\kappa$-$\mu$ shadowed was analyzed in~\cite{lopezMartinez}.
$2)$ $\alpha$-$\eta$-$\kappa$-$\mu$: As pointed out in~\cite{yacoubmodel}, this is a rather complex fading model that encompasses virtually all the fading channel models proposed in the literature based on a power envelope formulation.  Such a model incorporates the relevant short-term propagation factors, viz., non-linearity of the medium, scattered waves, specular components, and multipath clustering. The corresponding PLS performance undergoes $\alpha$-$\eta$-$\kappa$-$\mu$ fading channels was studied in~\cite{Mathur}. $3)$ Fluctuating Two-Ray: In this channel model, the receiver signal can be expressed as a superposition of two dominant waves, plus additional waves associated with diffuse scattering. Also, a fluctuation in the amplitude of the dominant rays is assumed. This fact is due to blockage by obstacles or by various electromagnetic disturbances. The secrecy performance over Fluctuating Two-Ray and Multiple-Rays fading channels were investigated in~\cite{Zeng}, and~\cite{nrays}, respectively. $4)$ Fisher-Snedecor ${\mathcal{F}}$: In this composite fading proposed in~\cite{nrays}, the received signal is modeled by jointly combining the effects of shadowing and small-scale fading. The PLS analysis over Fisher-Snedecor ${\mathcal{F}}$ fading channels was introduced in~\cite{Kong}.

Other works that analyze PLS performance with different network topologies over generalized fading channels are: Transmit Antenna Selection/Maximal Ratio Combining (TAS/MRC)~\cite{Hasna}, multiple-input multiple-output (MIMO)~\cite{Chen}, Full-Duplex~\cite{Zhou}, and Cognitive Radio System~\cite{Koo, Zou}.
 
The goal of this work is to provide a comprehensive survey of PLS on enabling technologies for 5G. Firstly, the main PLS performance metrics are introduced, including secrecy capacity, secrecy outage probability (SOP), alternative secrecy outage formulation, fractional equivocation, average information leakage rate, intercept probability, and the probability of strictly positive secrecy capacity (SPSC). A brief background on these metrics is also provided. Next, a theoretical framework on the PLS techniques commonly used to improve the secrecy performance is revisited. Then, we review the basic concepts of emerging 5G technologies. In particular, we focus on the following: massive MIMO, millimeter-wave (mm-Wave) communications, heterogeneous networks (HetNets), non-orthogonal multiple access (NOMA), and full-duplex (FD). Subsequently, we summarize the latest PLS research advances on the aforementioned 5G technologies.

The remainder of this paper is organized as follows. Section~\ref{sec:1} presents some fundamentals for PLS and reviews the main secrecy performance metrics. The PLS techniques are introduced in Section~\ref{PLStecnicas}. Section~\ref{sec:5G} summarizes concepts of promising 5G technologies and presents the recent advances in PLS research on these key 5G technologies. Section~\ref{sec:challenges} presents some of the
open challenges in wireless security communications, and provides some concluding remarks.

\section{Fundamentals of Physical Layer Security}
\label{sec:1}
Here, we introduce essential concepts to understand PLS in wireless communications systems.
\subsection{General System Model}
\begin{figure}[H]
\centering 
\psfrag{A}[Bc][Bc][0.7]{\textcolor{blue}{Transmitter}}
\psfrag{B}[Bc][Bc][0.7]{Bob}
\psfrag{E}[Bc][Bc][0.7]{Eve}
\psfrag{U}[Bc][Bc][0.7]{$h_{\mathrm{AB}}$}
\psfrag{W}[Bc][Bc][0.7][-15]{$h_{\mathrm{AE}}$}
\psfrag{Main channel}[Bc][Bc][0.7]{Main channel} 
\psfrag{Wiretap channel}[Bc][Bc][0.7]{Wiretap channel} 
\includegraphics[width=0.7\linewidth]{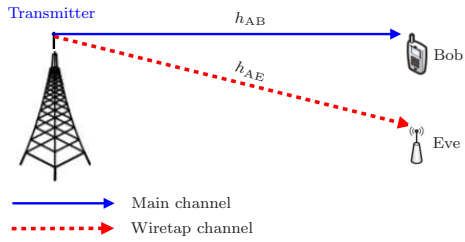} \caption{The wiretap channel model consisting of two legitimate nodes and an eavesdropper.}
\label{fig:sistema1}
\end{figure}

The general PLS model is made up of three main communication nodes as shown in Fig.~\ref{fig:sistema1}.

The first node is the legitimate transmitter (also known as Alice in network security jargon), the second node is the intended receiver (also known as
Bob), and the third node is the eavesdropper (also known as
Eve). The channel between Alice and Bob is known as the legitimate channel, while the path between Alice and Eve is named the wiretap channel (also known as
Eavesdropper channel). In this setup, Alice transmits confidential information to Bob, while Eve receives the signal and intends to decode it. Therefore, Alice's goal is to use a transmission approach that can deliver the secret information to Bob, while making sure that Eve cannot intercept the transmitted information. To attain secrecy in wireless systems, PLS uses signal processing techniques designed to take advantage of specific features of the channel including fading, noise, interference, among others.
Another relevant factor to take into account in the wiretap channel (see, Fig. 1) is the availability of channel state information
(CSI) in all the nodes (i.e., Alice, Bob, and Eve). CSI can vary from complete, partial
to even null at the nodes. From a secrecy perspective, CSI is of paramount importance because based on its knowledge,
the transmitter can decide whether or not to transmit and at which rate. Thus, this fact will lead to achieving a remarkable 
improvement on the SOP. However, in practice, all nodes can only obtain some kind of information about the channel between them and the other nodes. On the one hand, Alice is generally considered to know Bob's CSI but not Eve's CSI. This is because Eve is typically passive (i.e., Eve monitors the network, intercepts messages, and does not communicate with other users in the network). Several works such as~\cite{Labeau,Qing,Holger} have a done performance analysis of PLS with passive eavesdropper. On the other hand, there are scenarios in which Eve
is active and performs some of the following actions: intentional interference (also known as jamming), adulteration and modification or denial of service~\cite{Bhushan}. Performance analysis of PLS, which consider Alice knows Eve’s channel (i.e., active eavesdropper) can be found in~\cite{Liu,Timilsina,Petropulu}. 
It is worthwhile to mention that in the PLS evaluation, Eve’s and Bob’s channels are typically assumed to be independent of each other (i.e., both channels are separated at least half wavelength). Furthermore, the links (i.e., Alice-to-Bob and Alice-to-Eve) that do not meet the aforementioned condition (i.e., correlated channels) are investigated in~\cite{Le,peppas,Tsiftsis}.

\subsection{Performance Metrics}
 In this section, we explain the most used secrecy metrics proposed in the literature. Good knowledge of these metrics will ease the understanding of the works to be addressed in the following sections.

\subsubsection{Secrecy Capacity}
The secrecy capacity, $C_\mathrm{S}$, for a
wireless channel is the most used metric in PLS evaluation. $C_\mathrm{S}$ is defined as the capacity difference between the main and wiretap channels. Rigorously speaking, it defines the maximum secret rate at which the secret information reliably recovers at transmitter while remaining unrecoverable at Eve~\cite{Gopala}. Therefore, the $C_\mathrm{S}$ in a quasi-static fading channel case is formulated as in~\cite{Wyner} by 
\begin{align}\label{eq:8}
C_\mathrm{S}&=\!\text{max}\left \{ C_\mathrm{B}-C_\mathrm{E},0 \right \} \nonumber \\
&=\!\text{max}\left \{\mathrm{W}\log_2(1+\gamma_\mathrm{B})-\mathrm{W}\log_2(1+\gamma_\mathrm{E}),0  \right \}  
\end{align}
where $\abs{\cdot}$ is the absolute value, $\gamma_{\mathrm{X}}=\frac{|h_{\mathrm{AX}}|^2P_{\mathrm{A}}}{N_{\mathrm{0}}}$ for $\mathrm{X}$ $\in \left \{ \mathrm{B}, \mathrm{E} \right \}$ is the SNR, and $h_{\mathrm{AB}}$ and $h_{\mathrm{AE}}$ are the channel coefficients of the main and
wiretap channels, respectively. $P_{\mathrm{A}}$ is the transmit power at Alice, $N_{\mathrm{0}}$ is the average noise power, and  $C_\mathrm{B}$ and $C_\mathrm{E}$ are the capacities of the main and wiretap channels, respectively. Without loss of generality, it is considered a normalized bandwidth $\mathrm{W}=1$ in the aforementioned capacity formulations. Under this scenario, it is possible to attain secure transmissions only if the legitimate link has a better SNR than the eavesdropper link, i.e.,
\begin{align}\label{eq:9}
C_\mathrm{S}&=\left\{ 
        	\begin{array}{ll}
        		\hspace*{1mm} \log_2\left ( \frac{1+\gamma_\mathrm{B}}{1+\gamma_\mathrm{E}} \right ), \quad \text{if} \enspace \gamma_\mathrm{B}>\gamma_\mathrm{E}\\
        		\hspace*{1mm} 0, \hspace{6em} \text{if} \enspace \gamma_\mathrm{B} \leq \gamma_\mathrm{E},
        	\end{array}
        \right. \vspace{2mm} 
\end{align}
It is worthwhile to highlight that the $C_\mathrm{S}$ is widely extended by researchers to compute the SOP~\cite{Barros}.
\subsubsection{Secrecy Outage Probability}
The SOP is defined as the probability that the secrecy capacity falls below a target secrecy rate of $R_{\mathrm{S}}$. In other words, when the current $C_\mathrm{S}$ is not more than a pre-established target $R_{\mathrm{S}}$, the secrecy
outage happens. This fact means that the current secrecy rate cannot guarantee the security requirement. It can be formulated as in~\cite{Prabhu} by
 \begin{align}\label{eq:sop}
 \text{SOP}&=\Pr\left \{ C_\mathrm{S}\left ( \gamma_\mathrm{B},\gamma_\mathrm{E} \right ) < R_{\mathrm{S}}  \right \} 
\nonumber \\ 
&\stackrel{(a)}{=} \Pr\left \{ \left ( \frac{1+\gamma_\mathrm{B}}{1+\gamma_\mathrm{E}} \right ) < 2^{R_{\mathrm{S}}}  \right \} \nonumber \\
&\stackrel{(b)}{\geq} \Pr  \left \{ \frac{\gamma_\mathrm{B}}{\gamma_\mathrm{E}}< 2^{R_{\mathrm{S}}}\right \}
\end{align}
where $\Pr\left \{ \cdot  \right \}$ denotes probability. The SOP in~$(a)$ indicates that whenever $R_{\mathrm{S}}$ $<$ $C_\mathrm{S}$, the wiretap channel will be worse than the legitimate channel. So, secure communications are possible~\cite{Wang2018}. It is worth mentioning that the state of the art on the research topic of PLS over different types of fading channels focuses on the calculation of $(b)$ due to its simpler mathematical tractability concerning the formulation in $(a)$. Furthermore, the formulation in $(b)$ is well-known as the lower bound of the SOP and represents the ratio of two squared random variables (RVs), namely: $\gamma_\mathrm{B}$ and $\gamma_\mathrm{E}$, which can follow any fading distribution. In this context, to assess the PLS performance over generalized channels and their corresponding special cases, two recent works proposed in~\cite{Nogueira, Vega} developed closed-form fashions for the ratio of two squared RVs of the vast majority of fading channels models used to characterize the propagation environment of the 5G.
Despite the important insights that the SOP provides in the characterization of secrecy performance, it has the following demerits: $i)$ it cannot quantify the amount of data leaking to the eavesdroppers when the outage happens (i.e., transmission security); $ii)$ it cannot offer any information about the bob's skill to decode transmitted data successfully (i.e., transmission reliability); $iii)$ it cannot offer any information about the eavesdropper's skill to decrypt confidential data successfully; $iv)$ it cannot be straightly connected with quality of service (QoS) requirements for network services~\cite{Hamamreh}. Motivated by the limitations of the SOP, researchers in~\cite{Swindlehurst, Ronga} proposed new metrics to overcome the three aforementioned demerits of the SOP. Thus, the authors give more insights into PLS and how secrecy is measured. It is worthwhile to mention that the definition of the SOP and the $C_\mathrm{S}$ can also be used to the scenario with multiple antennas at different nodes. 
Readers are referred to~\cite{Rui, Long, Marcos} for further analysis of this field. Next, according to the classical SOP defined above, alternative secrecy outage formulations from~\eqref{eq:sop} are defined to follow.
\subsubsection{Alternative Secrecy Outage Formulation}
As previously mentioned the conventional SOP formulation in~\eqref{eq:sop} does not distinguish between reliability and security. Therefore, an outage event in~\eqref{eq:sop} can imply either a fault to achieve secrecy or that the transmitted message cannot be successfully decoded by Bob. In light of the above considerations, an alternative secrecy outage formulation was proposed in~\cite{Xiangyun}, which measures
that a transmitted data fails to attain secrecy. In a such formulation, the rate difference $R_\mathrm{E} \stackrel{\Delta}{=} R_\mathrm{B}-R_\mathrm{S}$ denotes the cost of security when the data is transmitted. Also, $R_\mathrm{B}$, is the rate of the transmitted messages, and $R_\mathrm{S}$ is the rate of the confidential data. It is worth mentioning that Bob can decode any transmitted message successfully if and only if $C_\mathrm{B}>R_\mathrm{B}$, whilst secrecy fails if $C_\mathrm{E}>R_\mathrm{E}$. Therefore, the alternative SOP can be formulated as the conditional probability upon a message being transmitted.
 \begin{align}\label{eq:sopalt}
 \text{SOP}_{\textrm{A}}=\Pr \left \{ C_\mathrm{E}>R_\mathrm{B} - R_\mathrm{S} | \textrm{message transmission} \right \}
\end{align}
Unlike the SOP definition in~\eqref{eq:sop}, the formulation in~\eqref{eq:sopalt} takes into consideration important system design parameters including the rate of the transmitted messages $R_\mathrm{B}$, and the fact whether a message was transmitted or not. Furthermore, this metric is useful when Alice knows the instantaneous Bob's CSI. Since in this scenario, Alice chooses whether or not to transmit, and if Alice decides to transmit, it will possibly do so with varying rates depending on Bob's CSI. In the contrast case, i.e., when the transmission is carried out at a constant rate\footnote{This scenario corresponds to the case when bob knows when Alice doesn't know about the instantaneous Bob's CSI}, the alternative SOP formulation in~\eqref{eq:sopalt} reduces to the unconditional probability.

This metric is achieving success in the latest research works related to performance in PLS. Readers can revise~\cite{Lima, HZhao, Cong, LiMu, LauHe} for more detailed information about this research topic.
\subsubsection{Fractional Equivocation Based Metrics}

Based on the limitation of the classic SOP in~\eqref{eq:sop} in measuring both the amount of data leakage to
the eavesdropper and Eve's skill to decode confidential data, three novel metrics was proposed in~\cite{He}. These metrics measure the secrecy performance of wireless systems from the partial secrecy perspective over quasi-static fading. The fractional equivocation (i.e., $\Delta$) is a random quantity due to the fading characteristics of the propagation medium. Mathematically, the fractional equivocation for a given fading realization of the channel is expressed as~\cite{He}

\begin{align}\label{eq:Deltau_1b}
\Delta=\left\lbrace 
	  \begin{array}{ll}
	1, & \mathrm{if}\ C_\mathrm{E} \leq C_\mathrm{B}-R_\mathrm{S} \\ 
	\left(C_\mathrm{B}-C_\mathrm{E}\right)/R_\mathrm{S}, & \mathrm{if}\ C_\mathrm{B}-R_\mathrm{S}< C_\mathrm{E} < C_\mathrm{B}\\
	0, & \mathrm{if}\ C_\mathrm{B}\leq C_\mathrm{E}.
	\end{array}  \right.
\end{align}
From~\eqref{eq:Deltau_1b}, the authors in~\cite{He} proposed the following metrics:\begin{enumerate}
  \item \textbf{Generalized Secrecy Outage Probability \\ (GSOP):} This metric is related to wireless systems with distinct secrecy levels measured in terms of Eve's capability to decode
the confidential information and is given by 
\begin{align}\label{eq:gsop}
\mathrm{GSOP}= \Pr\left \{ \Delta<\theta \right \},
\end{align}
where $0<\theta<1$ represents the minimum reasonable value of the fractional equivocation. Here, Eve's skill to decrypt the confidential message is set by selecting different values of $\theta$.
For instance, the conventional SOP is a particular case of the GSOP for $\theta=1$.
  \item \textbf{Asymptotic Lower
Bound on Eve's Decoding Error Probability:} This metric is defined as the average of the fractional equivocation and is given by
\begin{align}\label{eq:afefor}	
\bar \Delta =& \mathbb{E}\left[\Delta\right],
\end{align}
in which $\mathbb{E}\left[\cdot\right]$ is the expectation operation. It is worthwhile to mention that, when the entropy of data for transmission is long enough, Eve's decoding error probability for a given fading realization is lower bounded by the fractional equivocation, i.e., $P_e\geq\bar \Delta$\footnote{Interested readers can revise\cite[Eq. (6)]{He} for guidance about why the average fractional equivocation gives a lower bound of Eve's decoding error probability.}.
 \item \textbf{Average Information Leakage Rate:} This metric offers an idea of how fast the data is leaked to the Eve, when an unchanged rate transmission, $R_\mathrm{S}$, is adopted in the system\textcolor{red}{. It }can be expressed as 
 \begin{align}\label{eq:rl}
R_\mathrm{L}=\mathbb{E}\left[\left(1- \Delta\right) R_\mathrm{S}\right]=\left(1- \bar \Delta\right) R_\mathrm{S}.
\end{align}
\end{enumerate}
In~\cite{Phan,osorio}, researchers investigated the PLS performance by using different transmission topologies
based on the aforementioned metrics.

\subsubsection{Intercept Probability}
An intercept event happens when the $C_\mathrm{S}$ is negative or falls below 0. This means that the wiretap channel has a better SNR than the legitimate channel. The intercept probability can be formulated as in~\cite{Naeem} by
 \begin{align}\label{eq:pint}
 P_{\mathrm{int}} &=\Pr\left \{ C_\mathrm{S}\left ( \gamma_\mathrm{B},\gamma_\mathrm{E} \right ) < 0  \right \} \end{align}
Although this metric has not been widely explored in the literature, it is currently being investigated in evaluating the secrecy performance of wireless channels. Readers are referred to~\cite{Moualeu2,Choi,Jameel} for more information on this field of research.

\subsubsection{Probability of Strictly Positive Secrecy Capacity}
The Probability of SPSC is the
probability that the $C_\mathrm{S}$ remains higher than
0. This means that secrecy in communication has been attained\footnote{The authors in~\cite{Hamamreh} provide the theoretical meaning as well as the analytical expressions to quantify the $C_\mathrm{S}$ (e.g., perfect secrecy, weak secrecy, and strong secrecy) when the $C_\mathrm{S}$ is greater than zero.}. Mathematically, it can be written as in~\cite{Jameel2} by 
 \begin{align}\label{eq:Pspsc}
 P_{\mathrm{SPSC}} &=\Pr\left \{ C_\mathrm{S}\left ( \gamma_\mathrm{B},\gamma_\mathrm{E} \right ) > 0  \right \}. \end{align}
In~\cite{alpha,Kfading,Rician}, researchers investigated the security performance of wireless systems based on the SPSC metric over different fading channels models.

\section{Physical Layer Security Techniques}
\label{PLStecnicas}
This section introduces the background of PLS techniques commonly used in the research community.
\subsection{Artificial Noise Generation}
\begin{figure}[H]
\centering 
\psfrag{A}[Bc][Bc][0.7]{Alice}
\psfrag{B}[Bc][Bc][0.7]{Bob}
\psfrag{E}[Bc][Bc][0.7]{Eve}
\psfrag{U}[Bc][Bc][0.7]{Information}
\psfrag{W}[Bc][Bc][0.7][-23]{Artificial Noise}
\psfrag{Main channel}[Bc][Bc][0.7]{Main channel} 
\psfrag{Wiretap channel}[Bc][Bc][0.7]{Wiretap channel} 
\includegraphics[width=0.7\linewidth]{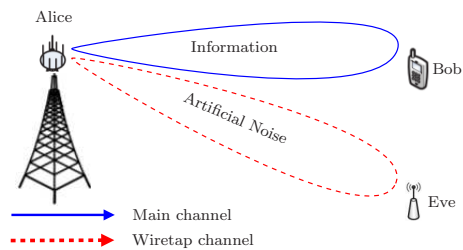} \caption{The model of AN network for a wiretap channel consisting of two main nodes and an eavesdropper.}
\label{AN}
\end{figure}

The main idea of this technique is to artificially degrade Eve's channel by injecting artificial noise (AN). The process consists in that an authorized node in the network (e.g., Alice, Bob, or another) adds well designed artificially jamming signals to the transmitted signal that can only harm Eve's channel~\cite{Cumanan}. The basic system model of AN network for PLS is depicted in Fig.~\ref{AN}.
In what follows, we review the fundamental works that use AN or jamming to improve PLS performance. In~\cite{Hayashi}, the authors proposed the design of AN-aided precoding to enhance PLS in a multi-user single eavesdropper wiretap visible light communication (VLC) networks. A fairness comparison of three AN-aided secure transmission approaches in wiretap channels was studied in~\cite{Malaney}. In such work, it is was demonstrated that regarding the secrecy performance the partially adaptive scheme (only the rate $R_\mathrm{B}$ changes) outperforms the on-off scheme (both $R_\mathrm{B}$ and confidential rate $R_\mathrm{S}$ vary). In~\cite{ANcite}, the researches proposed a $C_\mathrm{S}$ optimization (SCO)-AN to improve the $C_\mathrm{S}$ in wireless networks. The results in such an approach demonstrated that SCO-AN achieved greater improvement in the $C_\mathrm{S}$ than traditional AN.
\subsection{Multi-Antenna Diversity}
By leveraging the available spatial dimensions of wireless channels, MIMO techniques can diminish the impacts of fading while increasing the $C_\mathrm{S}$~\cite{wiley}. To achieve the full benefits of MIMO, the system must be protected against eavesdroppers attacks. 
\begin{figure}[H]
\centering 
\psfrag{A}[Bc][Bc][0.7]{Alice}
\psfrag{B}[Bc][Bc][0.7]{Bob}
\psfrag{E}[Bc][Bc][0.7]{Eve}
\psfrag{U}[Bc][Bc][0.7][23]{Main Channel}
\psfrag{W}[Bc][Bc][0.7][-15]{Wiretap Channel}
\psfrag{Main channel}[Bc][Bc][0.7]{Main channel} 
\psfrag{Wiretap channel}[Bc][Bc][0.7]{Wiretap channel} 
\includegraphics[width=0.7\linewidth]{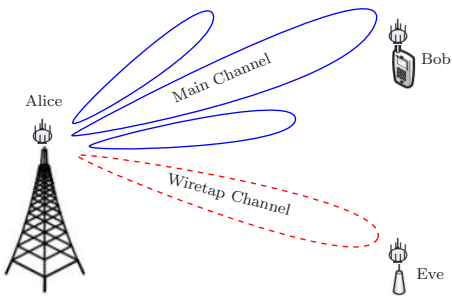} \caption{A MIMO wireless system by using secure beamforming with nulls directed towards Eve.}
\label{MultiAntena}
\end{figure}
In Multi-Antenna Diversity, the
basic idea of beamforming is to send the desired signal in the null space of the eavesdropper channel as shown in Fig.~\ref{MultiAntena}. A seminal work in~\cite{Mukherjee} was the first to investigate beamforming schemes for enhancing the PLS performance in MIMO wiretap channels. This paper encourages other researches to investigate beamforming challenges regarding PLS. Thus, in~\cite{Lv} the authors were the first to study PLS in a two-tier downlink HetNets. A novel layered PLS model was proposed in~\cite{BF3}. Here, the zero-forcing beamforming was applied to layered PLS  to tradeoff the achievable secrecy performance and the computational complexity. Moreover, an optimal technique commonly used on the receiver side in MIMO systems for improving PLS was presented in~\cite{Man}.

\subsection{Cooperative Diversity}
\begin{figure}[H]
\centering 
\psfrag{aa}[Bc][Bc][0.6]{a)}
\psfrag{bb}[Bc][Bc][0.6]{b)}
\psfrag{A}[Bc][Bc][0.6]{Alice}
\psfrag{B}[Bc][Bc][0.6]{Alice}
\psfrag{C}[Bc][Bc][0.6]{Bob}
\psfrag{D}[Bc][Bc][0.6]{Bob}
\psfrag{E}[Bc][Bc][0.6]{Eve}
\psfrag{F}[Bc][Bc][0.6]{Relay}
\psfrag{G}[Bc][Bc][0.6][16]{main channel}
\psfrag{H}[Bc][Bc][0.6][-20]{main channel}
\psfrag{BF}[Bc][Bc][0.6][-45]{wiretap channel}
\psfrag{mm}[Bc][Bc][0.6][20]{wiretap channel}
\psfrag{los}[Bc][Bc][0.6][19]{main channel}
\psfrag{lo}[Bc][Bc][0.6][-20]{main channel}
\psfrag{Small Cell}[Bc][Bc][0.8]{Untrusted Relay}
\includegraphics[width=0.7\linewidth]{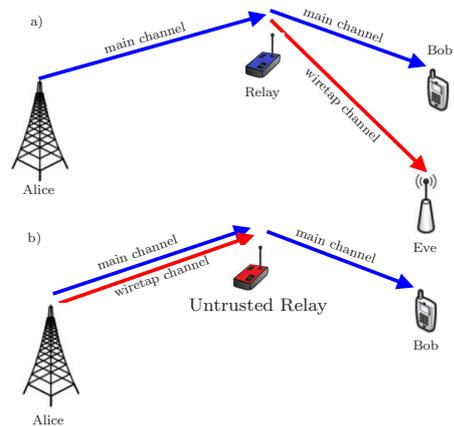} \caption{a) Traditional relay network in wiretap channel. b) Untrusted relay network in wiretap channel.}
\label{CD}
\end{figure}

In this section, we introduce cooperative communications, which besides providing reliability and extended coverage are used for improving the PLS performance. Relaying techniques allow the transmitter sends its information to the destination through a relay located between the two nodes. The most famous re-transmission protocols are: $i)$ amplify and forward (AF), and $ii)$ decode and forward (DF)~\cite{Hamamreh}. Relays can be configured in different ways to counteract eavesdropping. Specifically, they can behave like a conventional relay to attend the legitimate communication (vide Fig.~\ref{CD}a), or they can also act as jammers by sending AN to degrade Eve's channel. Moreover, they can take the role of potential eavesdroppers when they are untrusted. So, the confidential signals are vulnerable (vide Fig.~\ref{CD}b)~\cite{wiley}.
Next, we present interesting works on cooperative relaying methods to provide PLS in wireless systems. In~\cite{Poor}, the authors were the first to use cooperative relays for providing secure transmissions. In such work, three cooperative schemes were considered: DF, AF, and cooperative jamming (CJ)\footnote{In CJ, while the transmitter sends the data, the relay transmits an interference signal to harm the eavesdropper's channel.} to maximize the attainable secrecy rate subject to a transmit power constraint. 
The work in~\cite{Assi} studied the reachable secrecy diversity gain of cooperative networks with untrusted relays. In that approach, it was shown that the secrecy rate decreases as the number of untrusted relays increases. To enhance the PLS of the untrusted relay networks a new FD destination jamming (FDJ) topology was introduced in~\cite{Tan}. The results showed that FDJ strategies provided superior secrecy performance to that of the non-jamming schemes.

\section{Next Generation Physical Layer Technologies
}\label{sec:5G}
Next-generation cellular networks are planned to attain high capacity rates to face the rapid growth of data traffic.
The combination of 5G key technologies is considered as a cost-effective solution to cover the high QoS requirements in 5G. However, the dramatical increase in the amount of data and complex communication scenarios put forward higher requirements on the security of 5G. Here, we review the notions of each of the promising enabling technologies for 5G, including their advantages and disadvantages. Next, we summarize the latest research results of PLS for 5G technologies. 
\subsection{Massive MIMO}
\begin{figure}[H]
\centering 
\psfrag{A}[Bc][Bc][0.6]{$\text{U}_{\mathrm{1}}$}
\psfrag{B}[Bc][Bc][0.6]{$\text{U}_{\mathrm{2}}$}
\psfrag{C}[Bc][Bc][0.6]{$\text{U}_{\mathrm{k-1}}$}
\psfrag{D}[Bc][Bc][0.6]{Eve}
\psfrag{T}[Bc][Bc][0.8]{a) Downlink phase}
\psfrag{U}[Bc][Bc][0.8]{b) Uplink phase}
\psfrag{E}[Bc][Bc][0.6]{$\text{U}_{\mathrm{K}}$ }
\psfrag{S}[Bc][Bc][0.6]{$\text{N}_{\mathrm{
T}}$-antennas} 
\psfrag{Main Channel}[Bc][Bc][0.6]{Main channel} 
\psfrag{Wiretap Channel}[Bc][Bc][0.6]{Wiretap channel} 
\includegraphics[width=0.8\linewidth]{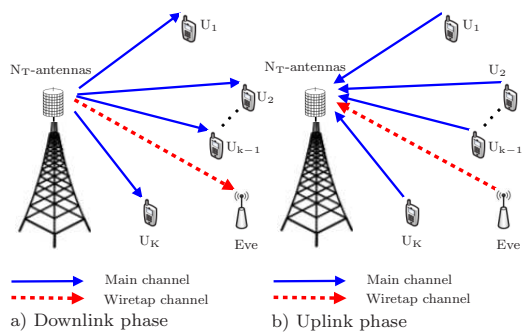} \caption{Massive MIMO downlink with $\mathrm{K}$ legitimate user nodes, $\text{U}_{\mathrm{k}}$ for $\mathrm{k} = 1,\cdots, \mathrm{K}$, and an eavesdropper.}
\label{massiveMimo}
\end{figure}

Massive MIMO is a multi-user topology in which the base station (BS) has a large number of antennas as depicted in Fig.~\ref{massiveMimo}. These arrangements provide several degrees of freedom for networks, better performance in channel capacities, and improve communication qualities in 5G networks~\cite{Schaefer}. For security purposes, massive MIMO gives a very oriented beam guides to the location of the legitimate user. So, the information leakage is reduced to undesired locations (i.e., Eve) significantly~\cite{Anwer}.
The authors in~\cite{Zhu2} were the first to investigate the drawbacks of PLS performance by assuming that the number of antennas goes to infinity (i.e., massive MIMO). Unlike the traditional MIMO, massive MIMO presents the following big challenges: $1)$  CSI estimation process is a difficult task; $2)$ the channels models are not independent as the distances of antennas are shorter than a half of the wavelength. Therefore, massive MIMO is still an open research field~\cite{masive}. Then, we survey the current security attacks of massive MIMO relying on passive and active eavesdropper cases, respectively.

\subsubsection{Passive Eavesdropper Scenarios}
The key concept here is that the existence of a passive eavesdropper does not affect at all the beam of transmission at the BS. So, it has a negligible effect on the $C_\mathrm{S}$. Recently, in~\cite{Xiao} an algorithm was developed to optimize power allocation of beam transmission for single-cell massive MIMO consisting of a passive eavesdropper with multiple antennas. The findings showed that beam transmission can attain optimal performance in terms of $C_\mathrm{S}$. Authors in~\cite{Guo10} investigated secure transmissions of multi-pair massive MIMO AF
relaying systems by considering Ricean fading. In that work, the attainable sum secrecy rate is maximized by using a power control topology. Also, the use of AN-aiding schemes to degrade the eavesdropping channel to improve the security in massive MIMO was analyzed in~\cite{Nguyen}. 

Other massive MIMO approaches with passive eavesdroppers include the effect of hardware deficiencies on the PLS performance of massive downlink MIMO in the existence of eavesdropper with multiple antennas~\cite{Derrick}, performance analysis of wireless communications in a multi-user massive MIMO by using imperfect CSI~\cite{LYang}, and SOP analysis for massive MIMO scenarios~\cite{Wei2}.  

\subsubsection{Active Eavesdropper Scenarios}
\begin{figure}[H]
\centering 
\psfrag{A}[Bc][Bc][0.6]{$\text{U}_{\mathrm{1}}$}
\psfrag{B}[Bc][Bc][0.6]{$\text{U}_{\mathrm{2}}$}
\psfrag{C}[Bc][Bc][0.6]{$\text{U}_{\mathrm{k-1}}$}
\psfrag{D}[Bc][Bc][0.6]{Eve}
\psfrag{Uplink With Pilot Contamination Attack}[Bc][Bc][0.8]{Uplink With Pilot Contamination Attac}
\psfrag{E}[Bc][Bc][0.6]{$\text{U}_{\mathrm{K}}$ }
\psfrag{S}[Bc][Bc][0.6]{$\text{N}_{\mathrm{
T}}$-antennas} 
\psfrag{Main Channel}[Bc][Bc][0.6]{Main channel} 
\psfrag{Wiretap Channel}[Bc][Bc][0.6]{Wiretap channel} 
\includegraphics[width=0.8\linewidth]{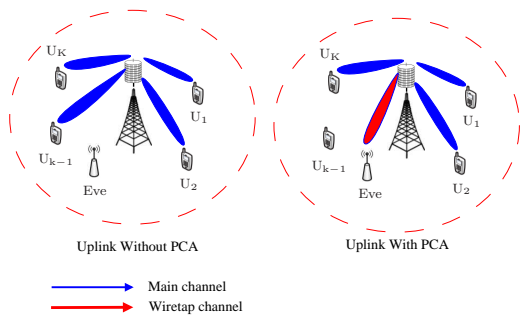} \caption{PCA on massive MIMO systems.}
\label{Pilot}
\end{figure}

A large number of PLS research works consider that Bob's CSI is known at Alice and does not take into account the process for obtaining this CSI. In time duplex division (TDD) massive MIMO, during the uplink phase, legitimate nodes transmit pilot signals to the BS to estimate the CSI for the later transmission of the downlink. At the same time, an active eavesdropper can interfere in the training stage to produce pilot contamination at the BS (see Fig.~\ref{Pilot}). This forces in the transmission phase (i.e., downlink) of the BS to inherently beamform towards the eavesdropper increasing its received signal power~\cite{Akgun}. This fact compromises that a secrecy rate may not be attainable. The result of this attack is that the advantages of PLS for massive MIMO are lost~\cite{MahamM}. To circumvent the referred limitation, the following works investigated techniques to avoid the pilot contamination attack (PCA). In~\cite{Hu},  
the authors proposed a reliable communication that does not need statistical information about the links for a TDD massive MIMO with an active eavesdropper. In the proposed transmission, an asynchronous protocol is used instead of the conventional synchronous protocol. A transmit power control policy was presented in~\cite{Kudathanthirige}, to allocate transmit power at the BS/relay for maximizing the attainable secrecy rate in Massive MIMO Downlink. For PLS in massive MIMO, in~\cite{GaoZhu} was designed robust scheme together with AN beamforming to deliver the legitimate nodes and eavesdroppers different signal-to-interference-and-noise ratio (SINR). 

Other secure massive transmissions against active eavesdropper include cooperative scheme strategy~\cite{YuanS}, data-aided secure downlink transmission scheme~\cite{Yongpeng},  and the secure communications design based on game theory~\cite{Sampaio}.

\subsection{mm-Wave}
\begin{figure}[H]
\centering 
\psfrag{A}[Bc][Bc][0.6]{$\text{U}_{\mathrm{1}}$}
\psfrag{B}[Bc][Bc][0.6]{$\text{U}_{\mathrm{2}}$}
\psfrag{C}[Bc][Bc][0.6]{$\text{U}_{\mathrm{3}}$}
\psfrag{E}[Bc][Bc][0.6]{$\text{U}_{\mathrm{4}}$}
\psfrag{F}[Bc][Bc][0.6]{$\text{U}_{\mathrm{5}}$}
\psfrag{G}[Bc][Bc][0.6]{$\text{U}_{\mathrm{k-1}}$}
\psfrag{H}[Bc][Bc][0.6]{$\text{U}_{\mathrm{K}}$ }
\psfrag{mm}[Bc][Bc][0.6][-35]{mm-Wave Backhaul}
\psfrag{los}[Bc][Bc][0.6][50]{LOS communication}
\psfrag{BF}[Bc][Bc][0.6][50]{Beamforming}
\psfrag{D}[Bc][Bc][0.6]{Eve}
\psfrag{Uplink Without Pilot Contamination Attack}[Bc][Bc][0.8]{plink Without Pilot Contamination Attack}
\psfrag{Uplink With Pilot Contamination Attack}[Bc][Bc][0.8]{Uplink With Pilot Contamination Attack}
\psfrag{S}[Bc][Bc][0.6]{$\text{N}_{\mathrm{
T}}$-antennas} 
\psfrag{Main Channel}[Bc][Bc][0.6]{Main channel} 
\psfrag{Wiretap Channel}[Bc][Bc][0.6]{Wiretap channel}
\includegraphics[width=0.8\linewidth]{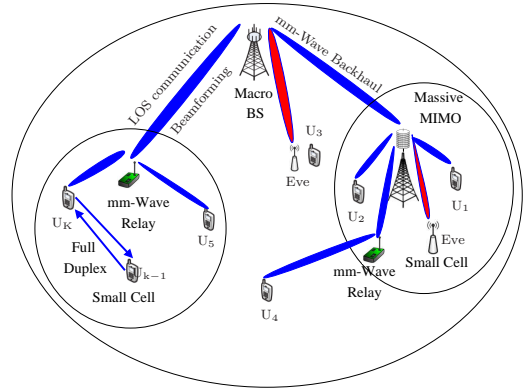} \caption{Illustration of promising technologies such as mm-Wave, massive MIMO, Full Duplex, and Small Cells.}
\label{5gTec}
\end{figure}

Nowadays, most wireless systems are allocated in the band spectrum of 300 MHz to 3 GHz, which is extremely full. In this context, mm-Wave\footnote{To have a more detailed framework about millimeter wireless systems, we refer the reader to~\cite{Heath}.} is a very innovative key solution for the next wireless networks (5G and beyond) to overcome this limitation. The idea behind mm-Wave communications is to take advantage of the unexploited high-frequency
mm-wave spectrum, ranging from 3-300 GHz to face with future multi-gigabit-per-second mobile applications. Unlike microwave networks, mm-Wave networks have several novel features, such as a large number of antennas\footnote{The small wavelength of high-frequency signals in mm-Wave enables a big number of antennas, which can be exploited to cover the requirements of massive MIMO. Therefore, the combination of massive MIMO, small cell geometries (which will be described later), and mm-Wave has vast potential to improve the security of the next networks~\cite{ZhengWang}.}, short-range, and different propagation laws~\cite{ChenWu}. The adoption of PLS mm-Wave networks systems is a remarkably emerging topic of research. Several approaches have been developed in this domain\footnote{For a good summary of works about the beginnings of PLS in mm-Wave, we refer the reader to the survey in~\cite{Khisti}}. The general model of PLS for mm-Wave, massive MIMO, FD, and Small Cells for 5G is presented in Fig.~\ref{5gTec}. Then, we review some of the current works to highlight the potential of this emerging field.
The major research papers focus on 28, 38, and 60 GHz band~\cite{Rappaport2}. 
In~\cite{JinXu}, to maximize the SNR (i.e., to improve the $C_\mathrm{S}$), the authors proposed AN aided two stages secure hybrid beamforming method in MIMO mm-Wave relay eavesdropping scenario. Here, the combination of the two-stage hybrid beamforming algorithm with AN allows guaranteeing both high throughput and communication security. The authors in~\cite{LeeJu} investigated secure communications techniques, namely, maximum ratio transmitting (MRT) beamforming, and AN beamforming. Specifically, it was developed the optimal power allocation between AN and the signal of interest that maximizes the $C_\mathrm{S}$ for AN beamforming. Concerning vehicular environments, in~\cite{Eltayeb}, the researchers proposed a location-based PLS technique for secure mm-Wave vehicular communication. Such a proposed technique takes advantage of a large number of antennas at the mm-Wave frequencies to jam eavesdroppers with sensitive receivers. The technique proved to offer good performance in terms of SOP.

Other approaches include PLS Analysis of Hybrid mm-Wave Networks~\cite{Satyanarayana}, and $C_\mathrm{S}$ of 5G mm-Wave Small Cells~\cite{Kadoch}.

\subsection{HetNets -- Small Cells}
\begin{figure}[H]
\centering 
\psfrag{A}[Bc][Bc][0.6]{$\text{U}_{\mathrm{1}}$}
\psfrag{B}[Bc][Bc][0.6]{$\text{U}_{\mathrm{2}}$}
\psfrag{C}[Bc][Bc][0.6]{$\text{U}_{\mathrm{3}}$}
\psfrag{E}[Bc][Bc][0.6]{$\text{U}_{\mathrm{4}}$}
\psfrag{F}[Bc][Bc][0.6]{$\text{U}_{\mathrm{5}}$}
\psfrag{G}[Bc][Bc][0.6]{$\text{U}_{\mathrm{6}}$}
\psfrag{H}[Bc][Bc][0.6]{$\text{U}_{\mathrm{7}}$}
\psfrag{I}[Bc][Bc][0.6]{$\text{U}_{\mathrm{8}}$}
\psfrag{J}[Bc][Bc][0.6]{$\text{U}_{\mathrm{9}}$}
\psfrag{K}[Bc][Bc][0.6]{$\text{U}_{\mathrm{10}}$}
\psfrag{L}[Bc][Bc][0.6]{$\text{U}_{\mathrm{k-1}}$}
\psfrag{M}[Bc][Bc][0.6]{$\text{U}_{\mathrm{K}}$ }
\psfrag{mm}[Bc][Bc][0.6][-35]{mm-Wave Backhaul}
\psfrag{los}[Bc][Bc][0.6][50]{LOS communication}
\psfrag{BF}[Bc][Bc][0.6][-50]{Beamforming}
\psfrag{D}[Bc][Bc][0.6]{Eve}
\psfrag{Uplink Without Pilot Contamination Attack}[Bc][Bc][0.8]{plink Without Pilot Contamination Attack}
\psfrag{Uplink With Pilot Contamination Attack}[Bc][Bc][0.8]{Uplink With Pilot Contamination Attack}
\psfrag{S}[Bc][Bc][0.6]{$\text{N}_{\mathrm{
T}}$-antennas} 
\psfrag{Main Channel}[Bc][Bc][0.6]{Main channel} 
\psfrag{Wiretap Channel}[Bc][Bc][0.6]{Wiretap channel}
\includegraphics[width=0.8\linewidth]{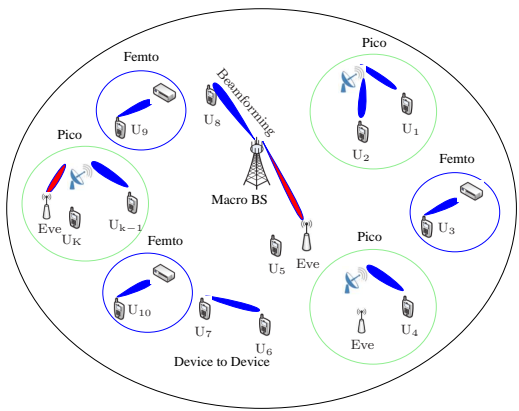} \caption{HetNets with legitimate users and eavesdroppers.}
\label{4tier}
\end{figure}

Traditionally, macro-cellular networks are efficient in offering area coverage for voice applications and services that support low data traffic but limited in providing high data rates. So, one of the promising solutions for users is to reduce the cell size in future wireless networks~\cite{Shafi}. In this context, HetNets will perform a pivotal role to meet the demands of 5G. The goal of HetNetsis to make efficient use of the spectrum to satisfy the spectacular growth of the data demands of the upcoming mobile services. In the HetNets topologies, users with different capabilities (i.e., transmission powers, coverage areas, etc.) are implemented to be part of a multi-tier hierarchical structure, as depicted in Fig.~\ref{4tier}. The high-power nodes (HPNs) with broad radio coverage fields are located in the macro cell, while low-power
nodes (LPNs) with limited coverage are located in small cells~\cite{Elkashlan}. The small cells (typically with coverage of a few meters) can have different configurations. For instance, the femtocells that are usually used in homes and development companies, and the picocells that are used for ample outdoor coverage~\cite{Shafi}. In addition, HetNets include a device level that incorporates device-to-device (D2D) communications. This technology favors nearby devices to connect directly and collaborate without using HPNs/LPNs, making them a strong tool for low-latency, and high-performance data services~\cite{D2D}.

The multi-tier topology in HetNets entails technical challenges (e.g., self-organization, backhauling, handover, and interference) to the investigation of PLS compared to the traditional single-tier architecture~\cite{Lopez}. Then, we review the most current works that address the aforementioned challenges of the HetNets on PLS. In two novel approaches~\cite{Schober,Schober2}, PLS performance in multi-cell networks has been studied. The researchers have taken advantage of cooperative multi-antenna transmissions to improve the $C_\mathrm{S}$
by assuming: $i)$ a single eavesdropper~\cite{Schober}, and $ii)$ a multiple untrusted relays~\cite{Schober2}. In~\cite{Haiyan}, the authors presented an interference-canceled opportunistic antenna selection (IC-OAS) topology to improve PLS in HetNets. Here, a passive eavesdropper is considered to intercept the communications of both the macro and small cells. 

Other secure communications works in HetNets systems include: Stochastic Geometry strategies~\cite{Kwok}, secrecy outage analysis undergo Nakagami-$m$ fading channels~\cite{Huadong}, and secure communications design based on game theory~\cite{Mingxuan}.

\subsection{Full-Duplex}
Among the promising technologies for 5G, FD technology carries major challenges for PLS communications. On one hand, FD enables the destination node to create AN to interfere with the eavesdropper and receive the data at the same time. On the other hand, if the eavesdropper has the FD technology, it can actively attack the receiver in the transmission while eavesdropping. Also, FD systems can double the spectral efficiency concerning the common half-duplex schemes. However, the main drawback that affects the transmission of FD is the management of the self-interference signal imposed by the transmission antenna on the receiving antenna within the same transceiver~\cite{Aghvami}. The research on FD PLS communication can be classified into four categorizations of FD PLS schemes\textcolor{red}{. Specifically,} the FD receiver, the FD transmitter and receiver, the FD BS, and the FD eavesdropper~\cite{Khisti}. Next, we review the most current works concerning the different configurations of the aforementioned FD technology.
In~\cite{Shihao}, the authors proposed a novel channel training (CT) method for a FD receiver to improve PLS. In this setup, the receiver (i.e., Bob) is equipped with $N_\mathrm{B}$ antennas. So, it can simultaneously receive the data and transmits AN to the eavesdropper. Here, to diminish the non-cancelable self-interference due to the transmitted AN, the destination node has to estimate the self-interference channel before the communication stage. In~\cite{Jongyeop} was considered a problem of a passive and clever eavesdropping attack on the MIMO wiretap scheme, where the receiver operates with FD mode. In such a system model, the clever eavesdropper cancels the interference (caused by the receiver) by stealing the CSI between legitimate nodes. To counteract this, the authors presented a cooperative jamming approach between transceivers to attain the optimal PLS performance. About FD active eavesdropper (FDAE), in~\cite{Junnan}, was analyzed the anti-eavesdropping and anti-jamming performance of D2D scenarios. In this case, the FDAE can passively intercept secret data in D2D topologies and actively jam all legitimate channels. In this respect, the authors proposed a hierarchical and power control method with multiple D2D node equipment and one cellular node to confront the smart FDAE.

Other works include FD strategies in HetNets~\cite{Babaei,Ahiadormey}, secrecy rate maximization in Wireless Multi-Hop FD Networks~\cite{Feng}, and secure communication based on joint design of information and AN beamforming for FD Networks~\cite{Yanjie}.
\subsection{Non-Orthogonal Multiple Access}
\begin{figure}[H]
\centering 
\psfrag{A}[Bc][Bc][0.6]{Base Station}
\psfrag{B}[Bc][Bc][0.6]{Internal Zone}
\psfrag{C}[Bc][Bc][0.6]{External Zone}
\psfrag{D}[Bc][Bc][0.6]{Eve}
\psfrag{Uplink Without Pilot Contamination Attack}[Bc][Bc][0.8]{plink Without Pilot Contamination Attack}
\psfrag{Uplink With Pilot Contamination Attack}[Bc][Bc][0.8]{Uplink With Pilot Contamination Attac}
\psfrag{E}[Bc][Bc][0.6]{User}
\psfrag{S}[Bc][Bc][0.6]{$\text{N}_{\mathrm{
T}}$-antennas} 
\psfrag{Main Channel}[Bc][Bc][0.6]{Main channel} 
\psfrag{Wiretap Channel}[Bc][Bc][0.6]{Wiretap channel} 
\includegraphics[width=0.8\linewidth]{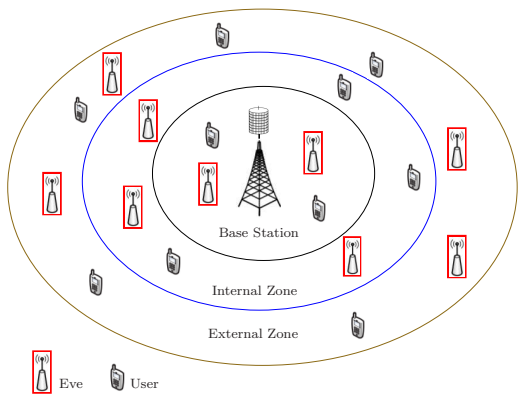} \caption{PLS model for NOMA}
\label{noma}
\end{figure}

Due to the limited spectral efficiency of orthogonal multiple access (OMA) systems in wireless networks, the OMA schemes are not appropriate to face the explosive growth in data traffic of the 5G. As a result, NOMA emerges as a promising candidate for 5G multiple access to provide massive connectivity and large system throughput~\cite{noma}. Furthermore, it is well-known that NOMA will use advanced reception techniques such as successive interference
cancellation (SIC) for robust multiple access. This fact may be a drawback in terms of processing delays. Fortunately, transmission/reception schemes in low-latency for NOMA systems
are being investigated in the literature. The basic NOMA model for PLS is shown in Fig.~\ref{noma}. There are two kinds of eavesdroppers scenarios: $i)$  the passive eavesdropper, whose channel
cannot be known at Alice; $ii)$ the active eavesdropper (i.e., common user), whose channel can be known
at Alice. Therefore, providing levels of security against the two types of eavesdroppers in NOMA technology is a challenging research topic in the design of the 5G networks~\cite{Hamamreh}. The main idea behind PLS for NOMA is to mitigate the security problems by finding the optimal power allocation policy that maximizes the secrecy sum rate (SSR) while satisfying the QoS requirements of users.

Then, we survey the key contributions regarding PLS in 5G NOMA systems. In~\cite{Ding}, the authors investigated the PLS performance in a single-input single-output (SISO) NOMA scheme by maximizing the SSR of the NOMA subject to the users’ QoS requirements. Here,  NOMA has proven a remarkable SSR improvement concerning the classical OMA. 
In~\cite{noma2}, the researchers proposed a secure transmission for downlink multiple-input single-output (MISO) NOMA  and energy-efficient design. In this approach, it was shown that the cooperative jamming NOMA scheme achieves much better secrecy performance than the direct transmission NOMA scheme. The secrecy in simultaneous wireless information and power transferring (SWIPT) in downlink NOMA systems was investigated in~\cite{Shojaeifard}. Later, the security challenges of vehicular users in an ultra-dense network were studied in~\cite{Chopra}. Here, it was demonstrated that NOMA-based multiple access is successful in attaining a high SSR for vehicular users by efficiently designing the allocation resources.

Other interesting works include PLS performance of uplink NOMA by using a stochastic geometry approach to analyze the effective secrecy throughput~\cite{Martinez}, the impact of random mobility on SSR maximization of NOMA systems subject to power limits and users' QoS requirements~\cite{Mobility}, the achievable secrecy rate by using the optimal security beamforming design in NOMA VLC networks~\cite{VLC}, and the SSR optimization for both primary users and randomly deployed secondary users in the NOMA underlay cognitive radio network~\cite{CRN}.

\section{Conclusions and Future Research Directions}
\label{sec:challenges}
This work has tackled the fundamentals concepts and techniques regarding PLS over the enabling 5G technologies. The following research topics emerge from the reviewed technologies in this survey:
  \begin{itemize}
         \item Accurate fading channel models play a remarkable role in an optimal secure transmission design over 5G. Thus, some efforts have been oriented to propose new more accurate channel models that provide a better fit to field measurements in a variety of new mm-Wave propagation scenarios. In this context, as claimed by the authors in~\cite{Paris} both Fluctuating Multiple-Ray and the $N$-Wave with Diffuse Power fading models constitute promising alternative models to characterize the propagation environment on mm-Wave communications. Therefore, the performance of PLS techniques over these generalized channels is an important topic for further investigations.
         \item Providing PLS usually entails compromising other system QoS requirements. For instance, high-security levels usually sacrifice throughput, while AN schemes compromise power efficiency. Based on these factors, the characterizing of the secrecy metrics in novel adversary models wireless through nontraditional (e.g., fractional equivocation, average information leakage rate, and GSOP) metrics are essential tracks in future research.  
         \item In the security paradigms, a promising direction of research is the integration of PLS and the classic wireless cryptography. Specifically, the physical layer features of the wireless medium can be exploited for designing new security algorithms to improve the current authentication and key management in higher layers. However, the integration of both approaches has not been properly studied at present\textcolor{red}{. Thus,} this topic needs further investigation.
         
         \item 
          An interesting future research direction could be to provide a detailed survey on the main drawbacks and merits of physical layer authentication (PLA) and Secret-Key Generation in 5G. In this sense, a research field, which is not yet investigated extensively in the literature, is the machine learning for intelligent PLA in 5G wireless networks.
          
            \item 
         Due to the combination of innovative technologies to cover the growing demands of data traffic and emerging services, it is essential to investigate PLS techniques regarding these new network scenarios. Within these networks, the following stand out: Unmanned Aerial Vehicles (UAV), enhanced Mobile Broadband (eMBB), Ultra-Reliable, and Low-Latency Communications (URLLC), massive Machine-Type Communications (mMTC), and Vehicle-to-Everything (V2X) networks.
\end{itemize}


\bibliographystyle{unsrt} 
\bibliography{references}   

%
%

\end{document}